\documentclass[showpacs]{revtex4} 
\usepackage{epsfig,amsmath,amssymb,amsbsy,graphicx}
\usepackage[cp1251]{inputenc}
\usepackage[english]{babel}  %

\begin{document}

\title{ THE EQUATION OF STATE AND THE QUASIPARTICLE MASS IN THE \\ %
DEGENERATE FERMI SYSTEM WITH AN EFFECTIVE INTERACTION}
\author{Yu.M.\,Poluektov}
\email{yuripoluektov@kipt.kharkov.ua} %
\affiliation{%
National Science Center ``Kharkov Institute of Physics and
Technology'', 1, Akademicheskaya St., 61108 Kharkov, Ukraine }%
\author{A.A.\,Soroka}
\affiliation{%
National Science Center ``Kharkov Institute of Physics and
Technology'', 1, Akademicheskaya St., 61108 Kharkov, Ukraine }%

\begin{abstract}
General formulas are derived for the quasiparticle effective mass
and the equation of state of the Fermi system with account of the
interparticle attraction at long distances and repulsion at short
distances. Calculations are carried out of the equation of state and
the effective mass of the Fermi system at zero temperature with the
use of the modified Morse potential. It is shown that the pair repulsive
forces promote a decrease of the effective mass, and the attractive
forces promote its increase. With a certain choice of the parameters of
the potential the dependence of the pressure on the density has a
nonmonotonic character, which enables to describe the coexistence of
the liquid and gaseous phases. The influence of three-body
interactions on the equation of state and the effective mass is
considered. The calculation results are compared with experimental
data concerning the quasiparticle effective mass in the liquid
$^3$He.
\newline%
\end{abstract}
\maketitle %

\section{INTRODUCTION}

Thermodynamic properties of the degenerate Fermi liquid are
determined by quasiparticles near the Fermi surface whose dispersion
law is of the same form as that of free particles, but with the
effective mass depending on thermodynamic quantities \cite{Landau,PN}. %
In the Fermi-liquid approach the effective mass is expressed through
a phenomenological amplitude that describes the interaction between
quasiparticles. On microscopic level the effective mass was
calculated for a weakly nonideal Fermi gas \cite{LP}. But such
calculations are true only for low-density systems, on the
assumption of validity of the Born approximation, when the
interaction between particles can be characterized by the only
parameter -- the scattering length that can be either positive or
negative, as in the case of $^3$He atoms.

In a more realistic description by means of model potentials, one
should take into account that the interaction between particles at
short distances has the strong repulsive character and at long
distances -- the attractive character. Therefore, in sufficiently
dense mediums the characteristics of a system depend substantially
on a concrete form of the interaction potential between particles
and its parameters, as well as on the density. At low particle
number density, when the average distance between particles is
comparable to or exceeds the characteristic range of action of the
potential, it is the long-range component of the interparticle potential
that plays a considerable role. With increasing the density and
decreasing the average distance between particles, the effects of
the strong repulsion between particles manifest themselves more and
more. In a simplified description of the interaction by means of the
only parameter, the scattering length, the effects connected with
the contributions of the long-range attraction and the short-range
repulsion cannot naturally be taken into account. Note also that the
Born approximation is not valid, as a rule, for the realistic
potentials with repulsion at short distances. It is known that
accounting for the strong repulsion (the hard core) at short
distances leads to considerable difficulties in calculations within
the self-consistent field model \cite{Jastrow,Thouless}. In
particular, the contribution of the short-range forces into a
system's energy appears to be considerably overestimated. In order to
bypass this difficulty, in this paper it is proposed to use the
effective interaction potential between particles differing from the
interaction potential between free particles, as it is adopted in
calculations by the self-consistent field method in nuclear physics
\cite{Skyrme,BBIG}. At that the ``hard'' core is replaced by the
``soft'' one, the value of which is an adjustable parameter.

The aim of this paper is to calculate the effective mass and to
derive the equation of state of the Fermi system in the microscopic
approach within a variant of the self-consistent field model that
was formulated in papers \cite{P1,P2}, by means of the interaction
potentials accounting for both the repulsion and attraction between
particles. A general formula is derived for the quasiparticle
effective mass in the Fermi system. The influence of the short-range
repulsive forces and the long-range forces of interparticle
attraction on the magnitude of the effective mass and the equation
of state is investigated. It is shown that the repulsive forces promote
a decrease of the effective mass, and the attractive forces promote its
increase. Specific calculations of the effective mass and the
equation of state are carried out at zero temperature with the use
of the modified Morse potential. With a certain choice of the
potential parameters, the dependence of the total pressure on the
density appears to be nonmonotonic and the equation of state
describes the coexistence of the gaseous and liquid phases. With
increasing the density the calculated effective mass of quasiparticles
decreases, which does not agree with experimental data on the liquid
$^3$He. It is suggested in the paper that a possible reason of
disagreement is connected with the effects of three-body forces in
the Fermi liquid. The role of three-body interactions is considered
and it is shown that accounting for three-body attractive forces
promotes an increase of the effective mass.

\section{The effective mass and the equation of state in the self-consistent field model}
In this section we will obtain general formulas for the
quasiparticle effective mass and the pressure. The wave function of
a quasiparticle in the Fermi system $\phi_j(q)$ within the framework
of the self-consistent field model in the formulation of papers
\cite{P1,P2} is obtained from the equation
\begin{equation} \label{EQ01}
\begin{array}{l}
\displaystyle{%
  \int\! dq' \big[ H(q,q')+W(q,q')\big]\phi_j(q')=\varepsilon_j \phi_j(q), %
}%
\end{array}
\end{equation}
where
$\displaystyle{H(q,q')=-\frac{\hbar^2}{2m}\Delta\,\delta(q-q')-\mu\,\delta(q-q')}$,
$m$ is the mass of a free fermion, $\mu$ is the chemical potential,
$q\equiv ({\bf r},\sigma)$ designates the spatial coordinate ${\bf
r}$ and the spin projection $\sigma$, $j\equiv (\nu,\sigma')$, where
$\nu$ is a full set of quantum numbers describing the state of a
particle except the spin projection $\sigma'$. We assume $s=1/2$. In
the formula (\ref{EQ01}) the self-consistent potential $W(q,q')$ has
the meaning of the mean field acting on a single particle \cite{P1,P2}. %
In the absence of an external field in the spatially uniform case
the state of a single particle can be characterized by its momentum
${\bf k}$, and its wave function has the form of a plane wave:
\begin{equation} \label{EQ02}
\begin{array}{l}
\displaystyle{%
  \phi_j(q)=\frac{\delta_{\sigma\sigma'}}{\sqrt{V}}\,e^{i{\bf k}{\bf r}}. %
}%
\end{array}
\end{equation}
Under these conditions and neglecting the magnetic effects, the
self-consistent potential has the form
$W(q,q')=\delta_{\sigma\sigma'}W({\bf r}-{\bf r}')$, where
\begin{equation} \label{EQ03}
\begin{array}{l}
\displaystyle{%
  W({\bf r})=W_0\delta({\bf r})+W_C(r). %
}%
\end{array}
\end{equation}
The form of the potentials $W_0, W_C(r)$, conditioned by the direct
and exchange interactions, is given below. In the spatially uniform
case from (\ref{EQ01}) it follows the expression for the
quasiparticle energy
\begin{equation} \label{EQ04}
\begin{array}{l}
\displaystyle{%
  \varepsilon_k=\frac{\hbar^2k^2}{2m} - \mu + W_0 + \frac{4\pi}{k}\int_0^{\!\infty}\!dr\,rW_C(r)\sin\!kr. %
}%
\end{array}
\end{equation}
At low but finite temperatures, we define the Fermi wave number for
the degenerate system by the relation
\begin{equation} \label{EQ05}
\begin{array}{l}
\displaystyle{%
  \frac{\hbar^2k_F^2}{2m} - \mu + W_0 + \frac{4\pi}{k_F}\int_0^{\!\infty}\!dr\,rW_C(r)\sin\!k_Fr =0. %
}%
\end{array}
\end{equation}
This relation establishes the connection between the chemical
potential and the Fermi wave number. With the help of the formula
(\ref{EQ05}), eliminating the chemical potential from (\ref{EQ04}),
we have
\begin{equation} \label{EQ06}
\begin{array}{l}
\displaystyle{%
  \varepsilon_k=\frac{\hbar^2k^2}{2m}-\frac{\hbar^2k_F^2}{2m} + 4\pi\int_0^{\!\infty}\!dr\,rW_C(r)\bigg( \frac{\sin\!kr}{k}-\frac{\sin\!k_Fr}{k_F} \bigg).   %
}%
\end{array}
\end{equation}
Near the Fermi surface $k=k_F+\Delta k$, and the effective mass is
defined by the relation
$\displaystyle{\varepsilon_k=\frac{\hbar^2}{m_*}k_F\Delta k=\frac{\hbar^2}{m_*}k_F(k-k_F) }$. %
It can be represented in the form
\begin{equation} \label{EQ07}
\begin{array}{l}
\displaystyle{%
  \frac{m}{m_*}=1 + \frac{2k_Fm}{\pi\hbar^2}\,J, %
}%
\end{array}
\end{equation}
where
\begin{equation} \label{EQ08}
\begin{array}{l}
\displaystyle{%
  J\equiv -\frac{2\pi^2}{k_F^2}\int_0^{\!\infty}\!dr\,r^3W_C(r)j_1(k_Fr), %
}%
\end{array}
\end{equation}
$j_1(x)=(\sin x-x\cos x)/x^2$ is the first order spherical Bessel
function. In this form the formula for the effective mass is true at
finite temperature and for an arbitrary character of the
interparticle interaction, and applicable for both the pair and
many-body forces. Note that the definition of the effective mass can
be given in somewhat different way, namely $m_*=\hbar k_F/v_F$
where $v_F=\partial\varepsilon_k/\hbar\partial k\big|_{k\,=k_F}$ \cite{LP}, %
but it leads to the same result (\ref{EQ07}).

The total pressure in the Fermi system with the pair interaction
in the self-consistent field model is determined by a sum of
two contributions $p=p_F+p_I$: the positive pressure of a gas of
quasiparticles with the effective mass
\begin{equation} \label{EQ09}
\begin{array}{ll}
\displaystyle{\hspace{0mm}%
  p_F= \frac{(3\pi^2)^{2/3}}{5}\frac{\hbar^2}{m_*}n^{5/3}, %
}%
\end{array}
\end{equation}
and the pressure conditioned by the pair interparticle interaction,
which can have both positive and negative sign:
\begin{equation} \label{EQ10}
\begin{array}{ll}
\displaystyle{\hspace{0mm}%
  p_I= 4\pi\int_0^{\!\infty}\!dr\,r^2U(r) \big(2\rho^2(0)-\rho^2(r) \big)=2\pi n^2\int_0^{\!\infty}\!dr\,r^2U(r)g(k_Fr), %
}%
\end{array}
\end{equation}
where $n$ is the particle number density, %
$\displaystyle{g(k_Fr)\equiv 1-\frac{9}{2}\bigg(\!\frac{j_1(k_Fr)}{k_Fr}\!\bigg)^{\!2} }$. %
Since the pair correlation function $g(k_Fr)$ is everywhere positive, then
the short-range positive part of the potential gives a positive contribution
to the pressure, and the long-range part gives a negative contribution.

\section{The effective potentials of the interparticle interaction}
Potential energy of a system of $N$ particles possessing an internal structure
can be represented as a sum of pair, three-body, etc. interactions
\begin{equation} \label{EQ11}
\begin{array}{l}
\displaystyle{%
  U({\bf r}_1,{\bf r}_2,\dots,{\bf r}_N)=\sum_{i<j}U({\bf r}_i,{\bf r}_j)+%
  \sum_{i<j<k}U({\bf r}_i,{\bf r}_j,{\bf r}_k)+\dots,
}%
\end{array}
\end{equation}
where $U({\bf r}_i,{\bf r}_j)=U({\bf r}_j,{\bf r}_i)$ , $U({\bf
r}_i,{\bf r}_j,{\bf r}_k)$  is a symmetric in all permutations of its
coordinates function. The self-consistent field theory with account of
contribution from three-body interactions is considered in paper \cite{P2}.

Let us discuss the question of choosing the potentials of interparticle
interactions. We begin with the analysis of the pair potentials.
Model potentials are often used to describe the interparticle interaction,
which in the simplest case depend only on the distance between particles.
An example of such potentials is given by the frequently used Lennard-Jones potential
\begin{equation} \label{EQ12}
\begin{array}{ll}
\displaystyle{%
  U_{L\!J}(r)= 4\varepsilon\!\left[\left(\frac{\sigma}{r}\right)^{\!12}-\left(\frac{\sigma}{r}\right)^{\!6}\right], %
}%
\end{array}
\end{equation}
containing two parameters: the distance $\sigma$ and the energy $\varepsilon$.
This potential rapidly tends to infinity at small distances.
It turns into zero at $r_*=\sigma$, becoming negative at $r>r_*$.
The potential (\ref{EQ12}) and similar ones can be written in the general form
\begin{equation} \label{EQ13}
\begin{array}{ll}
\displaystyle{%
 U(r)=\left\{
               \begin{array}{l}
                 U_{C}(r),\quad r<r_*, \vspace{2mm} \\
                 U_{L}(r),\quad r>r_*,
               \end{array} \right.
}%
\end{array}
\end{equation}
where $U_{C}(r)>0$ is the short-range part of the potential for which
it is in many cases assumed $U_{C}(r)\rightarrow\infty$ at $r\rightarrow 0$,
and $U_{L}(r)<0$ is the long-range part of the potential such that
$U_{L}(r)\rightarrow 0$ at $r\rightarrow\infty$.
The use of the potentials with the hard core in quantum mechanical calculations,
as it is known \cite{Jastrow,Thouless}, leads to considerable difficulties.
The potentials with the infinite core do not have the Fourier image and
the self-consistent field becomes infinite when using such potentials.
Sometimes, this fact is used as an argument to justify non-applicability of
the self-consistent field model in one or another case,
for example for describing the liquid \cite{Bazarov}.
The use of the potential which rapidly tends to infinity at small distances
means that an atom or other compound particle retains its identity
at arbitrary high pressures. Meanwhile, it is evident that
the critical pressure must exist at which atoms approach each other so closely
that they will be ``crushed'' and loose their identity. Therefore,
the requirement of absolute impermeability of particles at arbitrary
high pressures is excessively rigorous and unphysical and, in our opinion,
it is more reasonable to use the potentials which tend to a finite value
at small distances. An example of such potential is the known Morse potential
\begin{equation} \label{EQ14}
\begin{array}{ll}
\displaystyle{%
  U_M(r)=\varepsilon\big\{\!\exp[-2(r-r_0)/a]-2\exp[-(r-r_0)/a]\big\},
}%
\end{array}
\end{equation}
where $\varepsilon$ is the energy parameter and $r_0,a$ are specific distances.
It should also be noted that quantum chemical calculations lead to the potentials
having a finite value of energy at zero \cite{AS,ATB}. This energy
proves to be rather large. Thus, for the helium-helium potential it is of the order of $10^6\,$K.

But even at a finite but large energy, the role of the effects
conditioned by the hard core appears to be considerably overestimated.
It is associated with the fact that if, as it usually happens,
a wave function with a slowly varying in space modulus is used for describing a quasiparticle state
then the hard core gives an essentially overestimated contribution to the energy
and other system characteristics, to which attention was long ago called by Jastrow \cite{Jastrow,Thouless}.
Indeed, the presence of the hard core does not allow particles to approach each other
to distances less than its radius so that a real wave function,
as distinct from a plane wave for example, must decrease rapidly at such distances.
In order to remove the noted shortcoming, Jastrow \cite{Jastrow} suggested
a wave function that takes into account this circumstance,
though it appeared quite hard to use such function in calculations.

One can try to correct the situation in another way, namely,
having retained the description of quasiparticles by means of plane waves,
to reject the use of the ``realistic'' potentials by means of which free particles interact.
Instead of the ``hard'' potentials with a large value of the repulsive energy
at short distances, it is possible to use the ``soft'' effective potentials
for which the energy at short distances is a phenomenological adjustable parameter
determined by comparing some measurable quantity with an experiment.
It is this idea that was laid in the basis of calculations of nuclei
by the self-consistent field method \cite{Skyrme,BBIG}.
In these calculations, in order to bypass the problems arising due to
the strong repulsive core \cite{Brueckner},
the effective interaction is introduced without connection
to the basic interaction of free nucleons  \cite{Skyrme,BBIG}.

The simplest way of modification of the potentials (\ref{EQ13}) is
the replacement of their short-range part by a constant, so that
\begin{equation} \label{EQ15}
\begin{array}{ll}
\displaystyle{%
 \tilde{U}(r)=\left\{
               \begin{array}{l}
                 U_m,\qquad r<r_*, \vspace{2mm} \\
                 U_{L}(r),\quad r>r_*,
               \end{array} \right.
}%
\end{array}
\end{equation}
where $U_m>0$. Some of the known model potentials can be chosen as the long-range part.
Thus, for the Sutherland potential
\begin{equation} \label{EQ16}
\begin{array}{ll}
\displaystyle{%
 U_{L}(r)= -\varepsilon \big(\sigma/r\big)^6.
}%
\end{array}
\end{equation}
More realistic is the Kihara potential
\begin{equation} \label{EQ17}
\begin{array}{ll}
\displaystyle{%
  U_{L}(r)= 4\varepsilon\!\left[\left(\frac{\sigma}{r-a}\right)^{\!12}-\left(\frac{\sigma}{r-a}\right)^{\!6}\right], %
}%
\end{array}
\end{equation}
which contains an additional parameter of the length dimension
as compared with the Lennard-Jones potential, and turns into the latter at $a=0$ (\ref{EQ12}).

In numerical calculations in this paper we will use the Morse potential (\ref{EQ14}) as the long-range part.
Although the Morse potential is finite at all distances, its energy value at zero is still large
so that the positive part of this potential, according to the above said, will be replaced by a constant
thereby introducing an additional adjustable parameter of the length dimension.

As regards to the choice of the potential for three-body interactions,
very little is known about these potentials.
Note that the derivation from first principles of the potential of interaction of three atoms
as structureless entities presents a complex quantum mechanical problem \cite{Sarry}.
A review of the present state of the problem of three-body forces is given in paper \cite{Hammer}.

In particular, the three-body potential can be chosen in the form proposed in Ref. \cite{SS}:
\begin{equation} \label{EQ18}
\begin{array}{ll}
\displaystyle{%
  U(|{\bf r}-{\bf r}'|,|{\bf r}-{\bf r}''|)=u_0\,\exp\!\bigg[-\frac{\,|{\bf r}-{\bf r}'|+|{\bf r}-{\bf r}''|}{r_0}\bigg]. %
}%
\end{array}
\end{equation}
It is also possible to choose the potential in the form of the Gauss function:
\begin{equation} \label{EQ19}
\begin{array}{ll}
\displaystyle{%
  U(|{\bf r}-{\bf r}'|,|{\bf r}-{\bf r}''|)=\frac{u_0}{\pi^{3/2}r_0^3}\,\exp\!\left[-\frac{\,({\bf r}-{\bf r}')^2+({\bf r}-{\bf r}'')^2}{r_0^2}\right]. %
}%
\end{array}
\end{equation}
Such choice is characteristic in that in the limit $r_0\rightarrow 0$
the potential (\ref{EQ19}) turns into the three-body potential of zero radius which,
as shown in paper \cite{P2}, gives no contribution into the self-consistent field.
In principle, the model three-body potential can be chosen to depend on the three distances between three particles
\begin{equation} \label{EQ20}
\begin{array}{ll}
\displaystyle{%
  U({\bf r},{\bf r}',{\bf r}'')= U_3(|{\bf r}-{\bf r}'|,|{\bf r}-{\bf r}''|,|{\bf r}'-{\bf r}''|). %
}%
\end{array}
\end{equation}
In this paper we will use in calculations the potential just of this kind,
which is taken in the form of ``semi-transparent sphere'' potential:
\begin{equation} \label{EQ21}
\begin{array}{ll}
\displaystyle{%
 U_3(r,r',r'')=\left\{
               \begin{array}{l}
                 U_{3m},\quad r<r_3,\,r'<r_3,\,r''<r_3, \vspace{2mm}\\
                 \,\,0,   \qquad \textrm{else}.
               \end{array} \right.
}%
\end{array}
\end{equation}
The potential (\ref{EQ21}) is nonzero only in the case when
distances between each pair of three particles are less than $r_3$.

Both pair and three-body interactions give contribution into the self-consistent potential (\ref{EQ03})
$W_0=W_0^{(2)}+W_0^{(3)}$, $W_C(r)=W_C^{(2)}\!(r)+W_C^{(3)}\!(r)$. In the case of pair forces
\begin{equation} \label{EQ22}
\begin{array}{ll}
\displaystyle{%
 W_0^{(2)}=nU_0,\quad
 W_C^{(2)}\!(r)=-U(r)\rho(r).
}%
\end{array}
\end{equation}
Here $U(r)$ is the pair interaction potential, $U_0=\int U(r)d{\bf r}$. %
The one-particle density matrix has the form
\begin{equation} \label{EQ23}
\begin{array}{l}
\displaystyle{%
  \rho(r)=\frac{1}{2\pi^2r}\int_0^{\!\infty}\!\!dk\,k\sin(kr) f(\varepsilon_k), %
}%
\end{array}
\end{equation}
where $f(\varepsilon_k)=\big(\exp\beta\varepsilon_k +1\big)^{-1}$, and
the particle number density is $n=2\rho(0)$.

The contribution of three-body forces is determined by the relations \cite{P2}:
\begin{equation} \label{EQ24}
\begin{array}{l}
\displaystyle{\hspace{0mm}%
   W_0^{(3)}=2b_1-b_2,\quad W_C^{(3)}\!(r)=a_1\!(r)-2\rho(r)a_2\!(r).
}%
\end{array}
\end{equation}
For the three-body potential depending on distances between particles (\ref{EQ20}):
\begin{equation} \label{EQ25}
\begin{array}{ll}
\displaystyle{%
 a_1\!(r)=4\pi\sum_{l=0}^\infty\frac{1}{2l+1}  \int_0^{\!\infty}\!\! \underbar{\it{U}}_{\,3l}(r,r')\rho_l(r,r')\rho(r')r'^2dr', \qquad %
 a_2\!(r)=4\pi\rho(0)\!\int_0^{\!\infty}\!\! \underbar{\it{U}}_{\,30}(r,r')r'^2dr',  %
}\vspace{2mm}\\%
\displaystyle{\hspace{0mm}%
  b_1=16\pi^2\rho^2(0) \int_0^{\!\infty}\!\! r^2dr \!\int_0^{\!\infty}\!\! \underbar{\it{U}}_{\,30}(r,r')r'^2dr', \qquad %
  b_2=16\pi^2 \int_0^{\!\infty}\!\! r^2dr \!\int_0^{\!\infty}\!\! \underbar{\it{U}}_{\,30}(r,r')\rho^2(r')r'^2dr'. %
}%
\end{array}
\end{equation}
Here
\begin{equation} \label{EQ26}
\begin{array}{ll}
\displaystyle{\hspace{0mm}%
  U_3(r,r',|{\bf r}-{\bf r}'|)=\sum_{l=0}^\infty \underbar{\it{U}}_{\,3l}(r,r')P_l(\cos\theta), %
}\vspace{2mm}\\%
\displaystyle{\hspace{0mm}%
 \underbar{\it{U}}_{\,3l}(r,r')=\frac{2l+1}{2}\int_{-1}^{1}U_3\!\left(r,r',\!\sqrt{r^2+r'^2-2rr'x}\,\right)\!P_l(x)\,dx,%
}%
\end{array}
\end{equation}
\begin{equation} \label{EQ27}
\begin{array}{ll}
\displaystyle{\hspace{0mm}%
  \rho(|{\bf r}-{\bf r}'|)=\sum_{l=0}^\infty \rho_l(r,r')P_l(\cos\theta), %
}\vspace{2mm}\\%
\displaystyle{\hspace{0mm}%
  \rho_l(r,r')=\frac{2l+1}{2\pi^2}\int_0^{\!\infty}\!\!f(\varepsilon_k)j_l(kr)j_l(kr')k^2dk. %
}%
\end{array}
\end{equation}

\section{The effective mass, equation of state and chemical potential for the modified Morse potential}
In this section we will give the results of calculations of dependencies
of the effective mass, pressure and chemical potential on the particle number density at zero temperature,
for the effective pair potential of type (\ref{EQ15}) with the Morse potential (\ref{EQ14}) chosen as the long-range part.
The results of calculations will be compared with experimental data on the liquid $^3$He \cite{Peshkov,Wheatley}
at temperature close to zero. Note that the normal Fermi system is considered in this paper,
while $^3$He passes into the superfluid state at low temperatures \cite{Wheatley}.
However, since the superfluid transition temperature $T_C$ is very low,
of the order of several millikelvins, and it is much less than the Fermi temperature,
then under the fulfilment of the condition $T_C<T\ll T_F$
the normal system can be considered with a sufficient accuracy, assuming it to be at zero temperature.

The choice of the Morse potential as the long-range part is convenient because,
in particular, it admits the derivation of analytical expressions in some cases.
Thus, the integral (\ref{EQ08}) determining the effective mass (\ref{EQ07})
is given in this case by the formula:
\begin{equation} \label{EQ28}
\begin{array}{l}
\displaystyle{%
  J=\frac{U_m}{k_F^3}B_2(k_Fr_*)+\frac{\varepsilon}{k_F^3}\bigg[e^{2\gamma}I_M\bigg(k_Fr_*,\frac{2\gamma}{k_Fr_0}\bigg)-2e^{\gamma}I_M\bigg(k_Fr_*,\frac{\gamma}{k_Fr_0}\bigg)\bigg]. %
}%
\end{array}
\end{equation}
Here the designations are used:
\begin{equation} \label{EQ29}
\begin{array}{l}
\displaystyle{%
  B_2(z)\equiv\int_0^{z}\!y^2j_1^2(y)\,dy,\quad I_M(z,\alpha)\equiv\int_z^{\infty}\!e^{-\alpha y}y^2j_1^2(y)\,dy, %
}%
\end{array}
\end{equation}
$\gamma=r_0/a$ and $r_*/r_0=1-\ln 2/\gamma$.
With account of the relation between the Fermi wave number and the density $n=k_F^3/3\pi^2$,
the formulas (\ref{EQ07}), (\ref{EQ08}), (\ref{EQ28}) determine the dependence
of the effective mass on the density, shown in figure 1 (curve 1).
The choice of adjustable parameters in the pair potential will be discussed below.
It is seen that at low densities the quasiparticle effective mass increases
with increasing the density, which is in qualitative agreement with the calculation
for the low-density Fermi gas \cite{LP}, and then decreases with the density.
Qualitatively this result can be understood on the ground of the general formula (\ref{EQ07}).
As one can see, the positive part of the potential describing the repulsion leads to
a decrease of the effective mass, and its attractive negative part gives a contribution
that increases the effective mass. With increasing the density,
the average distance between particles decreases and, therefore,
the role of the interparticle repulsion increases leading to a decrease of the effective mass.
The obtained result does not correspond to the experimentally observed dependence of
the effective mass on the pressure in the liquid $^3$He,
which increases with increasing the pressure \cite{Peshkov,Wheatley}.
A possible reason of disagreement will be discussed in the next section.
\vspace{-11mm} %
\begin{figure}[h!]
\centering %
\includegraphics[width = 0.7\columnwidth]{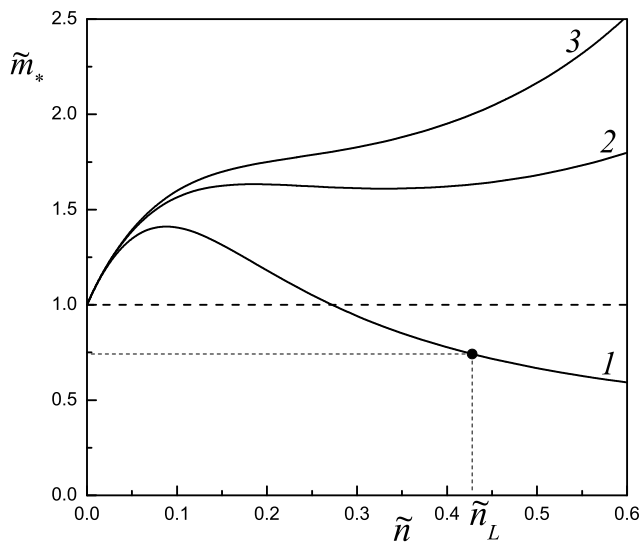} %
\vspace{-12mm}
\caption{\label{Fig01} %
Dependencies of the effective mass of quasiparticles of the Fermi liquid
$\tilde{m}_*=m_*/m$ on the dimensionless density $\tilde{n}=n/n_0$
($n_0=r_0^{-3}$). Curve {\it 1} is calculated
with the effective pair Morse potential and the parameters $U_m=46.55\,$K, $\gamma=4.16$.
The value $\tilde{n}_L=0.43$ corresponds to the density of the liquid $^3$He
at zero pressure and temperature, $\tilde{m}_*(\tilde{n}_L)=0.74$.
Curves {\it 2} and {\it 3} show dependencies of the effective mass with account of
both pair and three-body interactions for the parameters:
{\it 2} -- $U_{3m}=-4\varepsilon_0$, $r_3=1.27\,r_0$,
{\it 3} -- $U_{3m}=-4\varepsilon_0$, $r_3=1.30\,r_0$.
}%
\end{figure}

For the modified Morse potential, the contribution to the pressure
conditioned by the interaction is determined by the formula
\begin{equation} \label{EQ30}
\begin{array}{l}
\displaystyle{%
  p_I=\frac{2\pi n^2}{k_F^3}\bigg\{U_m I_{P1}(k_Fr_*) + \varepsilon\bigg[e^{2\gamma}I_{P2}\bigg(k_Fr_*,\frac{2\gamma}{k_Fr_0}\bigg)-2e^{\gamma}I_{P2}\bigg(k_Fr_*,\frac{\gamma}{k_Fr_0}\bigg)\bigg]\bigg\}, %
}%
\end{array}
\end{equation}
where
\begin{equation} \label{EQ31}
\begin{array}{l}
\displaystyle{%
  I_{P1}(z)\equiv\int_0^{z}\!g(y)y^2dy,\quad I_{P2}(z,\alpha)\equiv\int_z^{\infty}\!e^{-\alpha y}g(y)y^2dy,\quad  %
  \displaystyle{g(y)\equiv 1-\frac{9}{2}\bigg(\!\frac{j_1(y)}{y}\!\bigg)^{\!2} }. %
}%
\end{array}
\end{equation}
The dependence of the total pressure $p=p_F+p_I$ on the density is shown in figure 2. %
\vspace{-8mm} %
\begin{figure}[h!]
\centering %
\includegraphics[width = 0.7\columnwidth]{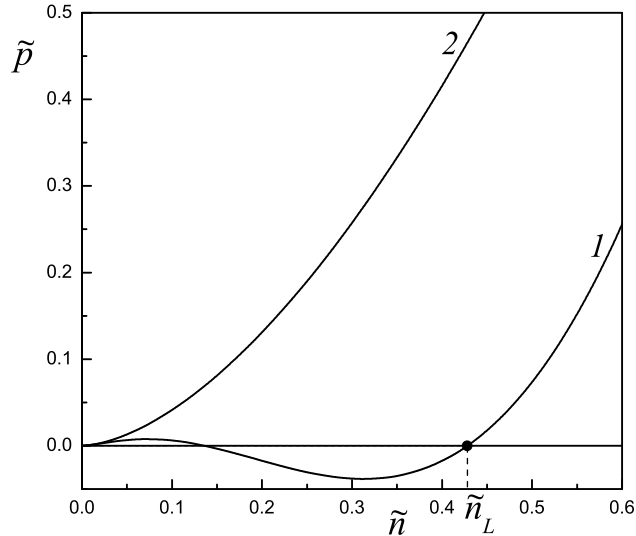} %
\vspace{-12mm}
\caption{\label{Fig02} %
Dependencies of the total pressure $\tilde{p}=p/p_0$
($p_0=\hbar^2/mr_0^5$) on the density $\tilde{n}$. Curve {\it 1}
corresponds to the Fermi liquid with the effective pair Morse potential
($U_m=46.55\,$K, $\gamma=4.16$), and curve {\it 2} -- the ideal Fermi gas.
}%
\end{figure}

As we see, the account for the interaction between atoms leads to a qualitative change
in the dependence of the pressure on the density as compared with the Fermi gas.
With chosen parameters of the interparticle interaction potential, the pressure appears
to be a nonmonotonic function of the density. There is a range of pressures in which
to each value of the pressure there correspond the two densities,
the larger of which matches the liquid and the lower matches the gas.
To obtain the equilibrium condition between phases, from the formula (\ref{EQ05})
we have to calculate the chemical potential which for the modified Morse potential has the form
\begin{equation} \label{EQ32}
\begin{array}{l}
\displaystyle{%
  \mu=\frac{\hbar^2k_F^2}{2m}+\frac{4U_m}{9\pi}\bigg[(k_Fr_*)^3-\frac{9}{2}I_{\mu 2}(k_Fr_*) \bigg] + %
}\vspace{2mm}\\%
\displaystyle{%
\hspace{4mm}+\frac{4\varepsilon e^{2\gamma}}{3\pi}\bigg[I_{\mu1}\bigg(k_Fr_*,\frac{2\gamma}{k_Fr_0}\bigg)-\frac{3}{2}I_{\mu3}\bigg(k_Fr_*,\frac{2\gamma}{k_Fr_0}\bigg)\bigg]- %
\frac{8\varepsilon e^{\gamma}}{3\pi}\bigg[I_{\mu1}\bigg(k_Fr_*,\frac{\gamma}{k_Fr_0}\bigg)-\frac{3}{2}I_{\mu3}\bigg(k_Fr_*,\frac{\gamma}{k_Fr_0}\bigg)\bigg]. %
}%
\end{array}
\end{equation}
Here the designations are used:
\begin{equation} \label{EQ33}
\begin{array}{l}
\displaystyle{%
  I_{\mu1}(z,\alpha)\equiv\int_z^{\infty}\!e^{-\alpha y}y^2dy,\quad I_{\mu2}(z)\equiv\int_0^{z}\!j_1(y)\sin(y)\, dy,\quad  %
  I_{\mu3}(z,\alpha)\equiv\int_z^{\infty}\!e^{-\alpha y}j_1(y)\sin(y)\,dy. %
}%
\end{array}
\end{equation}
The dependence of the chemical potential on the density is shown in figure 3.
In the ideal Fermi gas at zero temperature the chemical potential is positive
and coincides with the Fermi energy. The presence of the interaction between particles,
in particular the attraction at long distances, leads to a considerable change in
the dependence of the chemical potential on the particle number density as compared with
the case of the ideal gas. As seen (figure 3), this dependence can appear to be nonmonotonic
and the sign of the chemical potential to be negative.

Now we discuss the question of choosing the parameters of the effective potential.
The distance to the minimum and the depth of the well of the standard model potentials are known \cite{RS}.
For helium they are $\varepsilon=10.7\,$K, $r_0=2.97\,\textrm{{\AA}}$.
Generally speaking, when using the effective potential there is no reason to choose the values
of these parameters to be equal to the parameters of the interaction potential of free particles.
However, in numerical calculations in this paper we assume that $\varepsilon, r_0$
do not differ considerably from the above mentioned values.
There remain the two adjustable parameters $\gamma=r_0/a$ and $U_m$, which should be chosen
from additional considerations.
\begin{figure}[t!]
\vspace{-7mm}\centering %
\includegraphics[width = 0.7\columnwidth]{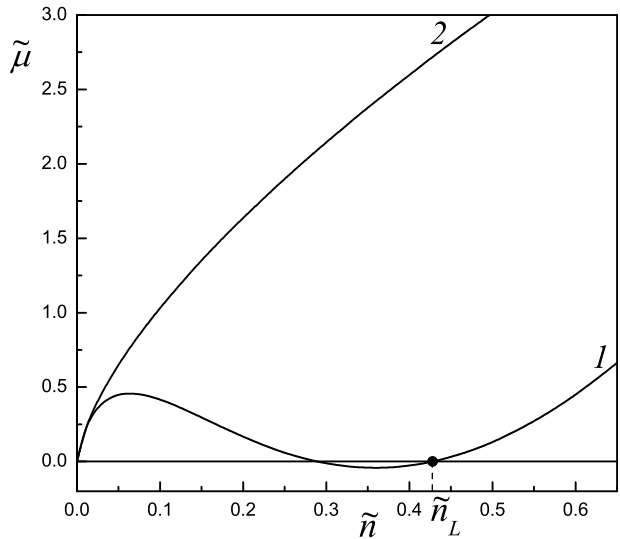} %
\vspace{-12mm}
\caption{\label{Fig03} %
Dependencies of the chemical potential $\tilde{\mu}=\mu/\varepsilon_0$
($\varepsilon_0=\hbar^2/mr_0^2$) on the density $\tilde{n}$. Curve {\it 1}
corresponds to the Fermi liquid with the effective pair Morse potential
($U_m=46.55\,$K, $\gamma=4.16$), and curve {\it 2} -- the ideal Fermi gas.
}%
\end{figure}

The character of the dependence of the total pressure on the density
depends on the choice of the parameter $\gamma$.
At a sufficiently large value of this parameter the pressure
monotonically increases with increasing the density.
With decreasing the noted parameter the dependence $p=p(n)$
becomes nonmonotonic, as shown in figure 2. Thermodynamically stable
are the regions on which $\partial p/\partial n>0$. There exists a range
of pressures to which there correspond the two values of the density,
the lower density matches the vapor and the larger matches the liquid.
It is known from the experiment \cite{Peshkov} that the pressure
of the saturated vapors of $^3$He tends to zero as temperature goes to zero.
Therefore, one of the conditions which should be satisfied
when choosing the adjustable parameters is the requirement
that the larger, different from zero density, at zero pressure
should coincide with the observed density of the liquid helium.

Since the equilibrium condition between phases is the equality of their
chemical potentials, then the second condition which should be satisfied is
the equality to zero of the liquid chemical potential, equalling
the vapor chemical potential as its density tends to zero.
Thus, the two adjustable parameters of the effective potential are determined
from the conditions
\begin{equation} \label{EQ34}
\begin{array}{l}
\displaystyle{%
  \mu(n_L;\gamma,U_m)=0, \quad p(n_L;\gamma,U_m)=0. %
}%
\end{array}
\end{equation}
With account of the fact that the density of the liquid $^3$He close to zero temperature
$n_L=1.635\!\times\!10^{22}\,\textrm{cm}^{-3}$, we find $\gamma=4.16, U_m=46.55\,$K.
Dependencies of the effective mass, pressure and chemical potential from the density
in figures {\it 1} -- {\it 3}\, are calculated for such values of these parameters.
It should be noted that the value of the adjustable parameter $U_m$ is considerably less
than the value of the Morse potential (\ref{EQ14}) at zero $U_M(0)\simeq 4\cdot\!10^4\,$K.

The effective mass $m_*/m\approx 0.74$ calculated for the indicated values of parameters
proves to be approximately four times less than the effective mass of quasiparticles
in the liquid $^3$He and, as shown in figure 1, decreases with increasing the density.
This does not agree with the observed rising dependence of the effective mass on the pressure
\cite{Peshkov,Wheatley}. The noted circumstance, apparently, is not connected with the choice
of the specific effective potential, since qualitatively similar dependencies arise also
when using the effective potentials that are different from the Morse potential used in this paper.
It can be suggested that the noted disagreement is connected with the contribution
from three-body interactions which role will be considered in the next section.

\section{The contribution from three-body interactions}
As a possible reason that explains an increase of the effective mass with
increasing the pressure and, consequently, the density, let us consider
the contribution from three-body interactions \cite{Hammer}.
The role of three-body interactions in the Fermi system is theoretically studied in paper \cite{P2}
where, in particular, it is shown that three-body forces give contribution into the self-consistent potential
only with account of their nonlocality. The contribution of three-body forces into the self-consistent potential
is given by the general formulas (\ref{EQ24}\,--\,\ref{EQ27}). Thus,
for the interaction potential of ``semi-transparent sphere'' type (\ref{EQ21}), we have:
\begin{equation} \label{EQ35}
\begin{array}{ll}
\displaystyle{\hspace{0mm}%
 \underbar{\it{U}}_{\,3l}(r,r')=U_{3m}\,\theta(r_3-r)\theta(r_3-r')(2l+1)
 \bigg[\theta(r_3-r-r')\delta_{l0}+\theta(r+r'-r_3)\frac{1}{2}\int_{x_0}^{1}\!P_l(x)dx \bigg],%
}%
\end{array}
\end{equation}
where $x_0=(r^2+r'^2-r_3^2)\big/2rr'$. At zero temperature
$f_k=\theta(k_F-k)$ and
\begin{equation} \label{EQ36}
\begin{array}{l}
\displaystyle{%
  \rho(r)=\frac{k_F^2}{2\pi^2r}j_1(k_Fr), \qquad
  \rho(0)=\frac{n}{2}=\frac{k_F^3}{6\pi^2}, \qquad
  \rho_l(r,r')=\frac{2l+1}{2\pi^2}\int_0^{k_F}\!\!dk\,k^2 j_l(kr)j_l(kr'). %
}%
\end{array}
\end{equation}
With account of the latter relations, we find
\begin{equation} \label{EQ37}
\begin{array}{ll}
\displaystyle{%
 a_1\!(r)=U_{3m}\,\theta(r_3-r)\frac{k_F^3}{2\pi^3}\,\frac{1}{k_Fr}\!\left[
          j_0^2(k_Fr_3)- j_0(k_Fr_3)j_0[k_F(r_3-r)] +
          \int_{\!-k_F(r_3-r)}^{k_Fr_3} dy\,j_0(k_Fr-y)j_1(y)  \right], %
}\vspace{2mm}\\%
\displaystyle{%
 a_2\!(r)=\frac{U_{3m}}{72\pi}(k_Fr_3)^3\,\theta(r_3-r) %
          \!\left[\left(\frac{r}{r_3}\right)^{\!3}-12\left(\frac{r}{r_3}\right) + 16 \right], %
}\vspace{2mm}\\%
\displaystyle{\hspace{0mm}%
  b_1=\frac{5\pi^2}{6}\rho^2(0)U_{3m}r_3^6 = \frac{5}{216\pi^2}U_{3m}(k_Fr_3)^6=0.0023\,U_{3m}(k_Fr_3)^6,  %
}\vspace{2mm}\\%
\displaystyle{\hspace{0mm}%
  b_2=\frac{U_{3m}}{12\pi^2}\Big[ %
      B_3(k_Fr_3)-12(k_Fr_3)^2B_1(k_Fr_3)+16(k_Fr_3)^3B_0(k_Fr_3) \Big], %
}%
\end{array}
\end{equation}
where $B_n(z)\equiv\int_0^{z}y^nj_1^2(y)dy$.
Dependencies of the effective mass on the particle number density with account of three-body forces,
calculated by the presented formulas, are shown on figure 1 (curves {\it 2}, {\it 3}).
The account for three-body forces with a negative interaction constant $U_{3m}<0$ leads to an increase of
the effective mass with increasing the density, which qualitatively agrees with
the experimentally observed dependence of the effective mass on the the pressure in $^3$He \cite{Peshkov,Wheatley}.
Note, however, that the speed of increase of the pressure with increasing the density slows down at that.
Although the account for the three-body forces of attraction gives the dependence of the effective mass on the density
that is more close to experiment, the problem of explaining the observed dependence on microscopic level
remains actual.

\section{Conclusion}
In this paper, on the basis of a variant of the self-consistent field theory
for Fermi systems that was formulated in papers \cite{P1,P2},
general formulas are derived for the quasiparticle effective mass and the equation of state.
The question is discussed about accounting for the strong repulsion between particles
at short distances (the hard core) in calculations within the self-consistent field model.
In view of the fact that the calculations with wave functions of the form of plane waves
do not account for the impossibility of approaching of particles to distances less than
the radius of the hard core, the contribution of the repulsion into all quantities
appears to be considerably overestimated in such calculations \cite{Jastrow}.
In order to bypass this difficulty, in this paper it is proposed to use in calculations
the effective potentials differing from the interaction potentials between free atoms,
similarly to how it is done in the self-consistent calculations of atomic nuclei \cite{Skyrme,BBIG}.
The long-range part of the effective potential describing the attraction is chosen to be the same as
in the usual model potentials, and the energy of repulsion is an adjustable parameter
which is found by comparison with experimental quantities.

The numerical calculations of the quasiparticle effective mass, pressure and chemical potential
at zero temperature are carried out for the effective potential with the long-range part
in the form of the Morse potential. It is shown that the repulsive forces promote a decrease of
the effective mass, and the attractive forces promote its increase.
With a certain choice of the parameters of the potential the dependence of the pressure on the density
has a nonmonotonic character, which enables to describe the coexistence of the liquid and gaseous phases.
The calculated effective mass of quasiparticles decreases with increasing the density,
which does not agree with the experimentally observed dependence in the liquid $^3$He.
It is suggested that the reason of the noted disagreement can be the effects conditioned by the action of three-body forces.
The influence of three-body interactions on the equation of state and the effective mass is considered,
and it is shown that in the case of attraction three-body forces lead to an increase of the effective mass.

\end{document}